\documentclass[aps,prl,twocolumn,showpacs,groupedaddress,floatfix]{revtex4}

\usepackage{graphicx}
\usepackage{amsmath,amssymb}
\usepackage{times,txfonts}

\begin{document}

\title{Role of Long-Range Correlations on the Quenching of Spectroscopic Factors}

\author{C. Barbieri}
\affiliation{Theoretical Nuclear Physics Laboratory, RIKEN Nishina Center, 2-1 Hirosawa, Wako, Saitama 351-0198 Japan}

\date{\today}

\begin{abstract}
  We consider the proton and neutron quasiparticle orbits around the closed-shell $^{56}$Ni and $^{48}$Ca isotopes.
It is found that large model spaces (beyond the capability of shell-model applications) are necessary for
 predicting the quenching of spectroscopic factors.
 The particle-vibration coupling is identified as the principal mechanism. Additional correlations---due
to configuration with several particle-hole excitations---are estimated using shell-model calculations
and generate an extra reduction which is $\alt$~4\% for most quasiparticle states.
The theoretical calculations nicely agree with ($e$,$e'p$) and inverse kinematics knockout experiments.
These results open a new path for a microscopic understanding of the shell-model.
\end{abstract}

\maketitle


{\em Introduction.}---
Single-particle states at the Fermi surface of shell closures
(or {\em quasiparticles)} play a crucial role in nuclear structure.
Their strengths [the {\em spectroscopic factors} (SFs)] and energies
are essential inputs in large-scale shell-model (SM) calculations,
which provide the most successful description of medium mass $0s1d$
and $0p1f$ nuclei~\cite{ANT.05}.
Single-particle wave functions and SFs also enter the nuclear
matrix elements that describe nucleon capture and emission
in stellar burning processes~\cite{Des.00}.

The need for understanding quasiparticles has become
more compelling in recent  years due to the realization that
their energies evolve, with changing proton or neutron number,
due to the nuclear tensor force~\cite{Sch.04,Ots.Ts}.
The quenchings of absolute spectroscopic factors (SFs) have also
been observed to change dramatically at the driplines~\cite{Gad.04,Gad.08}.
 These effects cause the breakdown of conventional magic numbers
far from the valley of stability and can give rise to new exotic
modes of excitation. 
In order to predict the properties of nuclei at the limits of stability,
it is imperative to derive robust models for quasiparticle properties.
Recently it has been suggested that {\em relative} SFs among small
quasiparticle fragments can be described  within
the SM approach~\cite{Tsa.09}.
This Letter discusses the problem of the {\em absolute quenching}
of SFs and shows that this requires accounting for coupling
to collective resonances and large model spaces, beyond
the capability of the SM.

The reactions for transfer of a nucleon to/from the initial state
$\vert \Psi^A_0 \rangle$ depend on the overlap wave function~\cite{pvbook,Han.03}
\begin{equation}
\psi^{A\pm1}_\alpha({\bf r}) ~=~ \langle \Psi^{A\pm1}_\alpha \vert \psi^{(\dag)}({\bf r}) \vert \Psi^A_0 \rangle \; ,
\label{eq:overlap}
\end{equation}
where $\alpha$ can label either particle or hole states.
SFs are identified with the normalization integral of $\psi^{A\pm1}_\alpha({\bf r})$
and give a ``measure'' of what fraction of the {\em final}
wave function, $\vert \Psi^{A\pm1}_\alpha \rangle$, can be
factorized into a (correlated) core plus an independent particle or hole.
Strong deviations from the independent particle model (IPM)---that is, a Slater
determinant with fully occupied orbits---signal substantial correlations
and imply the onset of non trivial many-body dynamics. 
For stable nuclei, a large body of data has been 
accumulated from ($e$,$e'p$) experiments.
These studies showed that proton absolute SFs for isotopes all across the nuclear chart
are {\em uniformly quenched} to 60-70\% of the IPM value~\cite{Lap.93,Kra.01}.
Short-range correlations (SRC)---originating from the repulsion of the
nuclear force at short distances---have been investigated as a possible cause.
However, they explain only a small fraction of the observed reduction~\cite{Pol.95,Nec.98},
as confirmed by recent experiments at Jefferson Laboratory~\cite{Roh.04,Sub.08}.
Thus, one is left to understand the role long-range correlations (LRC)
effects (see, for example, Refs.~\cite{Pan.97,Dic.04} and references therein).

Successful {\em ab-initio} calculations have been reported for
light nuclei using the variational Monte Carlo method~\cite{Lap.99}.
For larger masses, self-consistent (SC) Green's function (GF)
theory~\cite{Dic.04,dicvan} gave encouraging results for $^{48}$Ca
and $^{90}$Zr~\cite{Rij.90s}. However, these early calculations focused
only on one particular type of correlations [either coupling
to particle-hole (ph) or particle-particle and hole-hole phonons].
Recently, we have improved the SCGF method~\cite{Bar.01} to include correlations from 
the remaining collective modes, extended it to large-scale no-core calculations
up $^{56}$Ni, and based our studies on realistic nuclear forces~\cite{Bar.06,Bar.09c}.
%
Thus, we are now in the position of making accurate studies of the effects
of correlations.
In addition, the SCGF equations describe directly the interactions among
quasiparticle states, which allows extracting single-particle
orbits and effective interactions for the SM.
This Letter applies these new capabilities to study in detail quasiparticle states
in $^{56}$Ni and $^{48}$Ca and to extract an effective Hamiltonian for $0s1d$ and $0p1f$
orbits in the SM.
The SCGF results will then be compared to the SM ones to investigate the role
of {\em all} different types of correlations on the quenching of SF.

\begin{figure*}[t]
\includegraphics[width=0.28\columnwidth,clip=true]{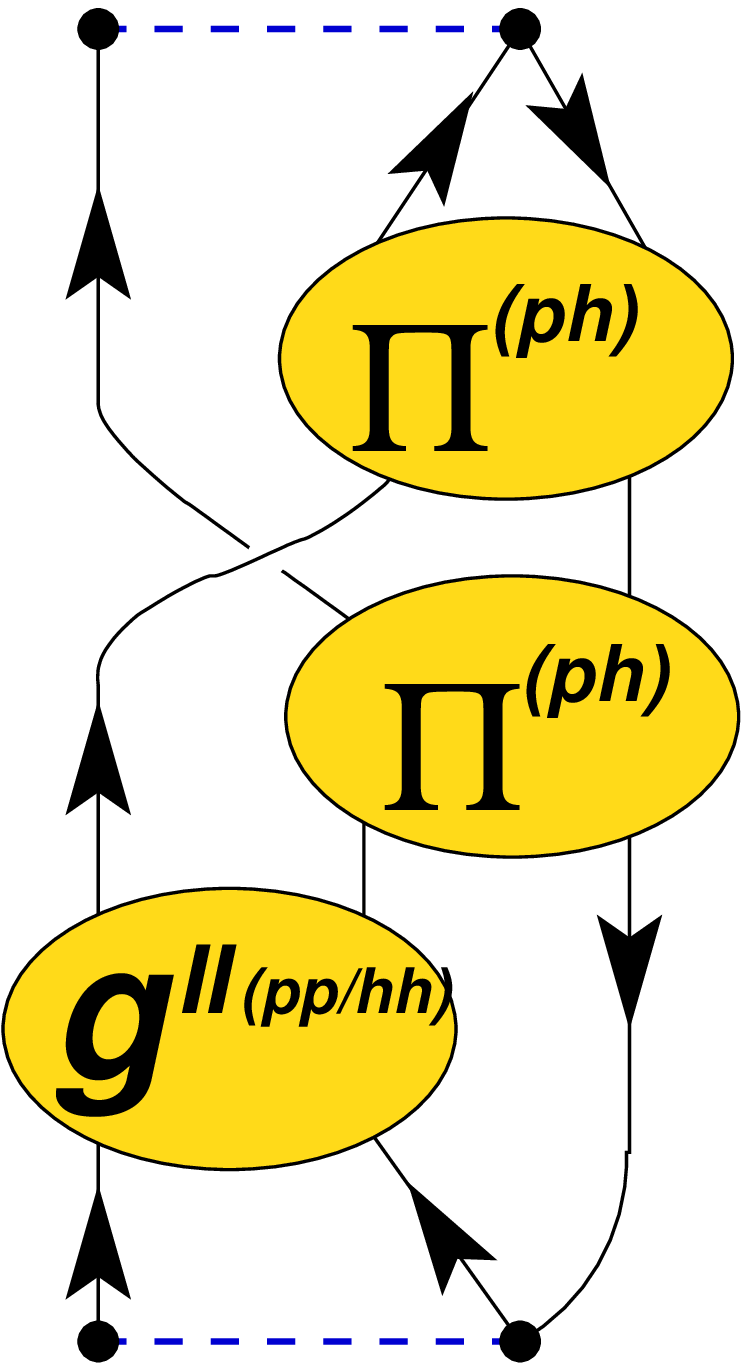}
\hspace{2cm}
\includegraphics[width=1.40\columnwidth,clip=true]{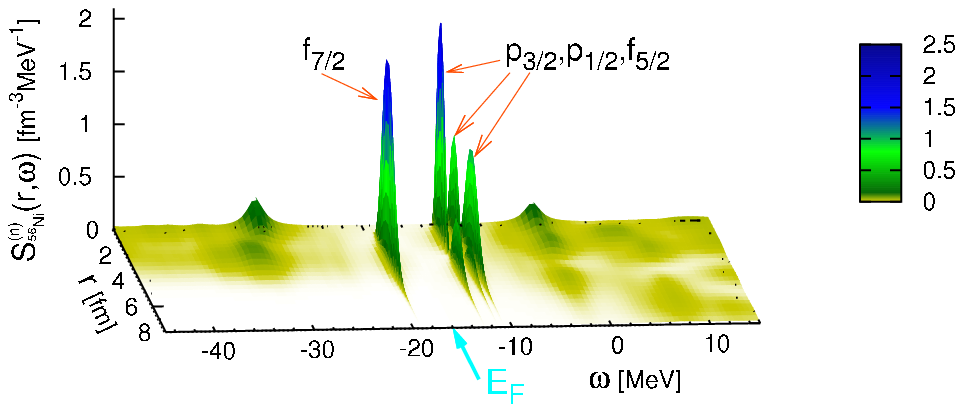}
\caption{(Color online)  {\em Left.}  One of the diagrams included in the 
correlated self-energy, $\tilde{\Sigma}(\omega)$. Arrows up (down) refer to
quasiparticle (quasihole) states, the $\Pi^{(ph)}$ propagators include collective ph
and charge-exchange resonances, and the $g^{II}$ include pairing between
two particles or two holes.   The FRPA method sums analogous diagrams,
with any numbers of phonons, to all orders~\cite{Bar.01,Bar.07}.
~%
{\em Right.}  Single-particle spectral distribution for neutrons in $^{56}$Ni,
obtained from FRPA. Energies above~(below) E$_F$ are for transitions to excited
states of $^{57}$Ni~($^{55}$Ni). The quasiparticle states close to the Fermi
surface are clearly visible. Integrating over {\bf r} [Eq.~(\ref{eq:sf})]
gives the SFs reported in Tab.~\ref{tab:Ni56}.
}
\label{fig:frpa}
\end{figure*}

{\em Formalism.}---
The calculations were performed in a large model space including of up to 10 major harmonic oscillator shells.
All details are reported in Ref.~\cite{Bar.09c} and here we give an overview of the points needed for
the present discussion.
 Spectroscopic factors were extracted from the one-body propagator
(spin/isospin indices are omitted for simplicity)~\cite{dicvan},
\begin{eqnarray}
g({\bf r}, {\bf r}';\omega) &=&
 \sum_n  \frac{ \left( \psi^{A+1}_n({\bf r}) \right)^* \;\psi^{A+1}_n({\bf r}') }
                       {\omega - (E^{A+1}_n - E^{A}_0) + i \eta } 
\nonumber \\
 &&~+~ \sum_k \frac{ \psi^{A-1}_k({\bf r}) \; \left( \psi^{A-1}_k({\bf r}') \right)^*  }
                       {\omega + (E^{A-1}_k - E^{A}_0) - i \eta } \; ,
\label{eq:g1} 
\end{eqnarray}
where the residues are the overlap amplitudes~(\ref{eq:overlap})
and the poles give the experimental energy transfer for nucleon pickup (knockout)
to the excited states of the systems with A+1 (A-1) particles. 
The propagator~(\ref{eq:g1}) is obtained by solving the Dyson equation 
[$g(\omega) ~=~ g^{(0)}(\omega) ~+~ g^{(0)}(\omega) \; \Sigma^\star(\omega) \; g(\omega)$],
where $g^{(0)}(\omega)$ propagates a free nucleon.
The information on nuclear structure is included in the irreducible
self-energy, which was split into two contributions:
\begin{equation}
\Sigma^\star({\bf r}, {\bf r}';\omega) = \Sigma^{MF}({\bf r}, {\bf r}';\omega) + \tilde{\Sigma}({\bf r}, {\bf r}';\omega) \; .
\label{eq:sigma}
\end{equation}
The term $\Sigma^{MF}(\omega)$ includes {\em both} the nuclear mean field (MF) and diagrams
describing two-particle scattering outside the model space, generated using a G-matrix
resummation~\cite{Jen.95}. As a consequence, it acquires
an energy dependence which is induced by SRC among nucleons~\cite{Bar.09c}.
The second term, $\tilde{\Sigma}(\omega)$, includes the LRC. In the present work,
$\tilde{\Sigma}(\omega)$ is calculated in the so-called Faddeev random phase
approximation (FRPA) of Refs.~\cite{Bar.01,Bar.07}.
This includes diagrams for particle-vibration coupling at all orders and with
all possible vibration modes, see Fig.~\ref{fig:frpa}, as well as
low-energy 2p1h/2h1p configurations.
Particle-vibration couplings play an important role in compressing the
single-particle spectrum at the Fermi energy to its experimental density.
However, a complete configuration mixing of states around the Fermi surface
is still missing and would require SM calculations.

Each spectroscopic amplitude $\psi^{A\pm 1}({\bf r})$ appearing in Eq.~(\ref{eq:g1})
has to be normalized to its respective SF as
\begin{equation}
Z_\alpha  = \int d{\bf r} \; \vert \psi^{A\pm 1}_\alpha ({\bf r}) \vert^2 = 
  \left.
  \frac{1}{  1   - 
  \frac{\partial \Sigma^\star_{\hat\alpha \hat\alpha}(\omega)}{\partial \omega}
  } \right\vert_{\omega=\pm(E^{A\pm 1}_\alpha - E^{A}_0)}
  \; ,
\label{eq:sf}
\end{equation}
where $\Sigma^\star_{\hat\alpha\hat\alpha}(\omega) \equiv <\hat{\psi}_\alpha|\Sigma^\star(\omega)|\hat{\psi}_\alpha>$
is the matrix element of the self-energy calculated for
the overlap function itself but normalized to unity
($\int d{\bf r} \; |\hat{\psi}_\alpha({\bf r})|^2=1$).
By inserting Eq.~(\ref{eq:sigma}) into~(\ref{eq:sf}), one distinguishes two contributions
to the quenching of SFs. 
For model spaces sufficiently large, all low-energy physics is described by $\tilde{\Sigma}(\omega)$.
Then, the derivative of $\Sigma^{MF}(\omega)$ accounts for the coupling to states outside
the model space and estimates the effects of SRC alone~%
\footnote{This approach includes explicitly the degrees of freedom outside the model space.
Hence, it does {\em not} require the renormalization of creation and annihilation operators
in Eq.~(\ref{eq:overlap}) that would be necessary for methods based on a similarity transformation.}.

In general, the SC self-energy (\ref{eq:sigma}) is a functional
of the one-body propagator itself, $\Sigma^\star=\Sigma^\star[g]$.
Hence the FRPA equations for the self-energy and the Dyson equation have
to be solved iteratively.
 The mean-field part, $\Sigma^{MF}[g]$, was calculated exactly in terms of 
the fully fragmented propagator~(\ref{eq:g1}).  For the FRPA, this procedure
was simplified by employing
the $\tilde{\Sigma}[g^{IPM}]$ obtained in terms of a MF-like propagator
\begin{equation}
 g^{IPM}({\bf r},{\bf r}';\omega) ~=~ 
 \sum_{n  \, / \hspace{-0.12cm} \in F}  \frac{ \left( \phi_n({\bf r}) \right)^* \;\phi_n({\bf r}') }
                       {\omega - \varepsilon^{IMP}_n + i \eta }  ~+~
 \sum_{k \in F} \frac{ \phi_k({\bf r}) \; \left( \phi_k({\bf r}') \right)^*  
}
                       {\omega - \varepsilon^{IMP}_k - i \eta } \; ,
\label{eq:gIPM}
\end{equation}
which is updated {\em at each iteration} to approximate Eq.~(\ref{eq:g1})
with a limited number of poles.
Eq.~(\ref{eq:gIPM}) defines a set of undressed single-particle states that
can be taken as a basis for SM applications. This feature will be used
below to estimate the importance of configuration mixing effects
on the quenching of spectroscopic factors.
 The present calculations employed the N3LO interaction from chiral perturbation
theory~\cite{Ent.03} with a modification of the tensor monopoles to correct
for missing three-nucleon interactions~\cite{Zuk.03}.

%
%
%
%

{\em Results.}---
The calculated single-particle spectral function
[\hbox{$S_{^{56}Ni}({\bf r},\omega) = \frac{1}{\pi}\vert g({\bf r}={\bf r}';\omega)\vert^2$}]
is shown in Fig.~\ref{fig:frpa} for the case of neutron transfer on $^{56}$Ni. 
This picture puts in evidence the quasiparticle and quasihole states associated with valence
orbits in the $0p1f$ shell. The corresponding SFs are reported
in Tab.~\ref{tab:Ni56}, including both protons and neutrons.
The first column is obtained by  including only the derivative of $\Sigma^{MF}(\omega)$
when calculating Eq.~(\ref{eq:sf}).
Since N3LO is rather soft compared to other realistic interactions the 
effect of SRC is relatively small.  From other models one could expect a quenching
up to about 10\%~\cite{Dic.04}, as confirmed by recent electron
scattering experiments~\cite{Roh.04,Bar.04,Sub.08}.
This difference would not  affect sensibly the conclusions below.
The complete FRPA result for SFs is given in the second column. 
For the transition between the $^{56}$Ni and $^{57}$Ni ground states, our result
agrees with knockout reactions with fast beams within the given error bar~\cite{Yur.06}.
Note that $^{56}$Ni is close to the stability line, where the analysis of these
reactions has been tested against electron scattering results.
The remaining orbits have similar quenchings and are in line with
results expected for stable nuclei~\cite{Lap.93}.
This shows that the dominant mechanism in reducing the spectroscopic
strength is to be looked for in LRC involving configurations among
several shells around the Fermi surface.

We now turn to the question of what are the relevant low-energy degrees of freedom
and the importance of configurations with multiple  ph excitations.
Excitations beyond 2p1h and 2h1p are partially included by the SC-FRPA, from
which one can expect an accurate description of quasiparticles around shell closures.
This was checked by comparing SC-FRPA and large-scale SM calculations
in the $1p0f$ shell.
 The basis and the interaction were constructed using the ``undressed'' single-particle orbits
$\phi_\alpha({\bf r})$, defined by Eq.~(\ref{eq:gIPM}), and the same G-matrix
interaction used in the full-space FRPA calculations. 
Shell-model calculations used the ANTOINE code~\cite{ANT.99,ANT.05} and included up to 6p6h excitations for the A=57 systems and
up to 8p8h for A=55. The resulting SFs are shown in the last three columns of Tab.~\ref{tab:Ni56}.
As an example, the quenching for the $p_{3/2}$ ground state of $^{57}$Ni, is predicted
to be 0.82 by FRPA and 0.79 by the SM.
Comparing with the full-space result of 0.65, one infers that
about half of the quenching is driven by degrees of freedom from outside the $1p0f$ space.
In particular, giant resonances originating from 1$\hbar\omega$ and 2$\hbar\omega$ excitations carry 
enough collectivity to influence the quenching of orbits close to the Fermi surface.
The difference between SM and FRPA in the $0p1f$ space
is $\Delta Z_\alpha$=0.79-0.82=-0.03,  which is smaller  than the typical model dependence
of SFs extracted from experiments.
One can treat the $\Delta Z_\alpha$s as small perturbations and add them to the full-space
FRPA results, leading to our final predictions for absolute SFs (third column of Tab.~\ref{tab:Ni56}).
%
%
Relative SFs with respect the other states in $^{57}$Ni have been studied
by re-analyzing the world ($d$,$p$) data~\cite{Tsa.09}. These results agree with the SM and, hence, with FRPA
ones (see Tab.~\ref{tab:Ni56}).  However, the model dependence of this analysis appears too large
to constrain absolute values of SFs.

\begin{table}[t]              
\begin{ruledtabular}
\begin{tabular}{lcccccccc}
                  &  \multicolumn{3}{c}{10 osc. shells}   & Exp.~\cite{Yur.06}  &~&  \multicolumn{3}{c}{$1p0f$ space}      \\
                  &   FRPA     &   full   &      FRPA         &            &~& FRPA   &   SM        & $\Delta Z_\alpha$ \\
                  &   (SRC)    &   FRPA   &+$\Delta Z_\alpha$ &            & &        &             &                   \\
\hline                 
$^{57}$Ni: \\
$\nu 1p_{1/2}$    &    0.96    &   0.63   &  0.61             &            & & 0.79  &   0.77     &  -0.02    \\
$\nu 0f_{5/2}$    &    0.95    &   0.59   &  0.55             &            & & 0.79  &   0.75     &  -0.04    \\
$\nu 1p_{3/2}$    &    0.95    &   0.65   &  0.62             & 0.58(11)   & & 0.82  &   0.79     &  -0.03    \\
$^{55}$Ni: \\                                                                                       
$\nu 0f_{7/2}$    &    0.95    &   0.72   &  0.69             &            & & 0.89  &   0.86     &  -0.03    \\
 \\
$^{57}$Cu: \\
$\pi 1p_{1/2}$    &    0.96    &   0.66   &  0.62             &            & & 0.80  &   0.76     &  -0.04    \\
$\pi 0f_{5/2}$    &    0.96    &   0.60   &  0.58             &            & & 0.80  &   0.78     &  -0.02    \\
$\pi 1p_{3/2}$    &    0.96    &   0.67   &  0.65             &            & & 0.81  &   0.79     &  -0.02    \\
$^{55}$Co: \\                                                                                       
$\pi 0f_{7/2}$    &    0.95    &   0.73   &  0.71             &            & & 0.89  &   0.87     &  -0.02    \\
\end{tabular}
\end{ruledtabular}
 \caption[]{Spectroscopic factors (given as a fraction of the IPM)
    for valence orbits around $^{56}$Ni.  For the SC-FRPA
    calculation in the large harmonic oscillator space, the values shown
    are obtained by including only SRC, SRC and LRC from particle-vibration
    couplings (full FRPA), and by SRC, particle-vibration couplings and extra
    correlations due to configuration mixing (FRPA+$\Delta Z_\alpha$).
     The last three columns give the results of SC-FRPA and SM in the restricted
    $1p0f$ model space. The $\Delta Z_\alpha$ are the differences between the last
    two results and are taken as corrections for the SM correlations
    that are not already included in the FRPA formalism.    
    }
\label{tab:Ni56}
\end{table}

%
%
%
%

Table~\ref{tab:Ca48} reports the results of the same analysis for $^{48}$Ca.
Also in this case, degrees of freedom from outside the first 10 oscillator shells have
a small effect, of 3-6\%, on the absolute SFs.
 The SC-FRPA coupling to collective modes reduces these to about 80\% of the IPM, and even
lower for proton holes to $^{47}$K.
The SM calculations employed the ``undressed'' single-particle orbits (\ref{eq:gIPM}) generated
by the large space SC-FRPA for this nucleus.
We adopted a model space (which will be named $sd_3pf$) consisting
of the ($1s_{1/2}$,$0d_{3/2}$,$1p_{3/2}$,$0f_{7/2}$) orbits for protons and of 
($1p_{1/2}$,$1p_{3/2}$,$0f_{5/2}$,$0f_{7/2}$) for neutrons. 
This can be handled by present day computers with up to 10p10h configurations
and includes the most important excitations across the Fermi surface.
The SM results in the $sd_3pf$ model space do not give additional quenchings for the valence orbits in the
$1p0f$ shell. Hence, multiple ph configurations are suppressed
even more than for the $^{56}$Ni case.
Note that the $\nu 0f_{5/2}$ quasiparticle in $^{49}$Ca splits in two main
fragments but their total strength remains unchanged.
Experimental SFs for the ground ($\nu 0p_{3/2}$)  and first excited ($\nu 0p_{1/2}$) states,
have been extracted from a combined analysis of ($d$,$p$) and ($n$,$\gamma$) reactions~\cite{Muk.08}.

Proton-hole states in $^{47}$K are a case in which SM correlations start becoming important.
The FRPA in the $sd_3pf$ space describes about a half of the quenching of the $0d_{3/2}$ orbits in the full space.
The SM calculations induce a further reduction of $\approx$8\%. 
Adding this estimate  brings the $0d_{3/2}$ SF
to 0.59, in agreement within the experimental ($e$,$e'p$) value of 0.56(5)~\cite{Kra.01}.
Shell-model correlations are even more substantial for the $1s_{1/2}$ fragments: here the $sd_3pf$ space
is sufficient to reproduce almost all of the quenching obtained in the largest space.
Configurations not included in the FRPA are also more important and induce a further reduction of 
up to 16\% for the ground state of $^{47}$K.
The total predicted strength of the two $1s_{1/2}$ peaks is 0.50, somewhat lower that the ($e$,$e'p$) value of 0.61(5).

Utsuno and collaborators have also calculated the spectral strength for
proton removal from $^{48}$Ca and found that the spin-orbit spitting between
the $0d$ states and the ordering of single-particle energies are sensitive
to the tensor force~\cite{Uts.09}.
These calculations also show that the fragmentation pattern in $^{47}$K
is well reproduced by the SM, as suggested in~\cite{Tsa.09}.
The experimental strength is however lower by an overall factor of $\approx$0.73.
This is nicely explained by the combined SC-FRPA and SM results, where the final
corrected strengths of Tab.\ref{tab:Ca48} are smaller than the pure SM ones by
0.75 for the $0d_{3/2}$ peak and by 0.70 for the summed $1s_{1/2}$ strength.

\begin{table}[t]                                       
\begin{ruledtabular}                                    
\begin{tabular}{lcccccccc}
                   &  \multicolumn{3}{c}{10 osc. shells}       &  Exp. & & \multicolumn{3}{c}{$sd_3pf$ space}  \\
                   & FRPA   &   full     &      FRPA           &                    &~& FRPA  &   SM    & $\Delta Z_\alpha$ \\
                   & (SRC)  &   FRPA     &  +$\Delta Z_\alpha$ &                    &~&       &         &                   \\
\hline                                                                                    
$^{49}$Ca: \\                                                                                                               
$\nu 0f_{5/2}$     &  0.97  &    0.10    &      0.16           &                    & &  0.19 &   0.25  &     +0.06  \\  
$\nu 0f_{5/2}{}^*$ &  0.97  &    0.69    &      0.62           &                    & &  0.77 &   0.70  &     -0.07  \\  
$\nu 1p_{1/2}$     &  0.97  &    0.82    &      0.82           & 0.71($^{+20}_{-12}$)& & 0.96 &   0.96  &   $<$0.01  \\  
$\nu 1p_{3/2}$     &  0.97  &    0.78    &      0.78           & 0.53(11)           & &  0.94 &   0.94  &   $<$0.01  \\  
$^{47}$Ca: \\                                                                                                              
$\nu 0f_{7/2}$     &  0.96  &    0.80    &      0.80           &                    & &  0.97 &   0.97  &   $<$0.01  \\  
 \\ 
$^{49}$Sc: \\                                                                                                                
$\pi 1p_{3/2}$     &  0.95  &    0.52    &      0.48           &                    & &  0.74 &   0.70  &     -0.04  \\  
$\pi 0f_{7/2}$     &  0.95  &    0.74    &      0.73           &                    & &  0.94 &   0.93  &     -0.01  \\  
$^{47}$K: \\                                                                                                             
$\pi 0d_{3/2}$     &  0.94  &    0.67    &      0.59           & 0.56(4)            & &  0.87 &   0.79  &     -0.08  \\  
$\pi 1s_{1/2}{}^*$ &  0.95  &    0.52    &      0.36           & 0.53(4)            & &  0.64 &   0.48  &     -0.16  \\  
$\pi 1s_{1/2}$     &  0.95  &    0.24    &      0.14           & 0.08(1)            & &  0.34 &   0.23  &     -0.10  \\  
\end{tabular}                                                          
\end{ruledtabular}                                                      
 \caption{Same as Tab.~\ref{tab:Ni56} but for $^{48}$Ca.
 The $\nu 0f_{5/2}$ particle ($\pi 1s_{1/2}$ hole) fragments in
 $^{49}$Ca ($^{47}$K) with lowest excitation energy are indicated
 by an asterisk.
  The experimental SFs for $^{49}$Ca are from Ref.~\cite{Muk.08}, those
 for $^{47}$K are form Ref.~\cite{Kra.01}.}
\label{tab:Ca48}
\end{table}

{\em Summary.}---
In conclusion, we have studied the effects of correlations on the quenching of
absolute spectroscopic factors. The present work does not support the assumption that 
the uniform quenching observed in stable nuclei originates from a unique mechanism such as
short-range correlations.
SRC have been predicted~\cite{Pol.95,Dic.04} and experimentally confirmed~\cite{Roh.04,Sub.08}
to influence just a small fraction of the total strength.
We find that spectroscopic factors for valence orbits result from a combination
of configuration mixing at the Fermi surface and coupling to collective resonances. The former
can usually describe the relative spectroscopic factors (hence, the fragmentation pattern at low energy)
and it will be crucial in open shell nuclei. On the other hand, the overall quenching of
absolute spectroscopic factors requires both the coupling to collective modes at higher energies and 
large model spaces that cannot be approached by 
shell-model calculations.
This work reported on the first consistent calculation of all these effects in the medium-mass
region and used the same realistic interaction.

It was also found that quasiparticles around shell closures can be described
accurately by self-consistent Green's function theory in the FRPA approximation. Moreover, the theory
links the interactions among quasiparticle states with the underlying nuclear forces.
This opens a new possibility for deriving single-particle orbits, effective charges and interactions
for the  shell-model form realistic nuclear interactions.

It is a pleasure to thank W.~H.~Dickhoff for several enlightening discussions.
This work was supported by the Japanese Ministry of Education, Science and Technology (MEXT) under KAKENHI grant no. 21740213.

\end{document}